\documentclass[twocolumn,traditabstract]{aa}  
\usepackage{graphicx}
\usepackage{txfonts}
\usepackage{amsmath,amsfonts,amssymb}
\usepackage{fixltx2e}
\usepackage{caption,subcaption}
\usepackage[breaklinks, colorlinks, citecolor=blue]{hyperref}
\usepackage{color}
\usepackage{natbib}
\bibpunct{(}{)}{;}{a}{}{,}

\def\setsymbol#1#2{\expandafter\def\csname #1\endcsname{#2}}
\def\getsymbol#1{\csname #1\endcsname}

\newbox\tablebox    \newdimen\tablewidth
\def\leaderfil{\leaders\hbox to 5pt{\hss.\hss}\hfil}
\def\endPlancktablewide{\tablewidth=\textwidth 
    $$\hss\copy\tablebox\hss$$
    \vskip-\lastskip\vskip -2pt}
\def\tablenote#1 #2\par{\begingroup \parindent=0.8em
    \abovedisplayshortskip=0pt\belowdisplayshortskip=0pt
    \noindent
    $$\hss\vbox{\hsize\tablewidth \hangindent=\parindent \hangafter=1 \noindent
    \hbox to \parindent{$^#1$\hss}\strut#2\strut\par}\hss$$
    \endgroup}
\def\doubleline{\vskip 3pt\hrule \vskip 1.5pt \hrule \vskip 5pt}

\newcommand{\Planck}{{\em Planck}}
\newcommand{\Msolar}{M_\odot}

\newcommand{\Rfive}{R_{500}}
\newcommand{\Mfive}{M_{500}}
\newcommand{\rhoo}{\rho_{\rm 0}}
\newcommand{\rs}{r_{\rm s}}
\newcommand{\cfive}{c_{\rm 500}}
\newcommand{\thetam}{\theta_{\rm m}}
\newcommand{\Dclus}{D_{\rm clus}}
\newcommand{\DCMB}{D_{\rm CMB}}
\newcommand{\Kappafiver}{K_{\rm 5\theta 500}}
\newcommand{\estKappafiver}{\hat{K}_{\rm 5\theta 500}}
\newcommand{\sigmaKappa}{\sigma_{\Kappafiver}}
\newcommand{\Mfiver}{M_{\rm 5R 500}}
\newcommand{\estMfiver}{\hat{M}_{\rm 5R 500}}
\newcommand{\estMfive}{\hat{M}_{500}}
\newcommand{\Mx}{M^{\rm X}_{500}}

\newcommand{\thetat}{\theta_{\rm t}}
\newcommand{\thetak}{\theta_{\rm k}}
\newcommand{\yo}{y_{\rm o}}	
\newcommand{\estyo}{\hat{y}_{\rm o}}
\newcommand{\jnu}{j_\nu}		
\newcommand{\Tth}{T_{\rm t}}
\newcommand{\Tk}{T_{\rm k}}
\newcommand{\tth}{t_{\rm t}}
\newcommand{\tk}{t_{\rm k}}

\newcommand{\nui}{\nu_i}
\newcommand{\Psit}{\Psi_{\thetat}}
\newcommand{\Phim}{\Phi_{\thetam}}
\newcommand{\sigt}{\sigma_{\thetat}}

\newcommand{\Sigmam}{\Sigma_{\rm m}}
\newcommand{\Sigmacr}{\Sigma_{\rm cr}}
\newcommand{\phio}{\phi_0}
\newcommand{\estphio}{\hat{\phi_0}}

\newcommand{\estphi}{\hat{\phi}}
\newcommand{\estphiks}{\hat{\phi}_{\rm ks}}
\newcommand{\estphikn}{\hat{\phi}_{\rm kn}}
\newcommand{\estphikk}{\hat{\phi}_{\rm kk}}

\begin{document}

   \title{Measuring cluster masses with CMB lensing: a statistical approach}

   \author{Jean-Baptiste Melin
         \inst{1}
         \and
         James G. Bartlett\inst{2}\fnmsep\inst{3}
         }

   \institute{DSM/Irfu/SPP, CEA-Saclay, F-91191 Gif-sur-Yvette Cedex, France\\
              \email{jean-baptiste.melin@cea.fr}
        \and
            APC, AstroParticule et Cosmologie, Universit\'e Paris Diderot, CNRS/IN2P3, CEA/lrfu, Observatoire de Paris, Sorbonne Paris Cit\'e, 10, rue Alice Domon et L\'eonie Duquet, 75205 Paris Cedex 13, France\\
            \email{bartlett@apc.univ-paris7.fr}
            \and
            Jet Propulsion Laboratory, California Institute of Technology, 4800 Oak Grove Drive, Pasadena, California, U.S.A.
            }

   \date{Received ; accepted }

 
  \abstract {We present a method for measuring the masses of galaxy clusters using the imprint of their gravitational lensing signal on the cosmic microwave background (CMB) temperature anisotropies.  The method first reconstructs the projected gravitational potential with a quadratic estimator and then applies a matched filter to extract cluster mass.  The approach is well-suited for statistical analyses that bin clusters according to other mass proxies.  We find that current experiments, such as \Planck, the South Pole Telescope and the Atacama Cosmology Telescope, can practically implement such a statistical methodology, and that future experiments will reach sensitivities sufficient for individual measurements of massive systems.  As illustration, we use simulations of \Planck\ observations to demonstrate that it is possible to constrain the mass scale of a set of 62 massive clusters with prior information from X-ray observations, similar to the published \Planck\  ESZ-XMM sample.  We examine the effect of the thermal (tSZ) and kinetic (kSZ) Sunyaev-Zeldovich (SZ) signals, finding that the impact of the kSZ remains small in this context.  The stronger tSZ signal, however, must be actively removed from the CMB maps by component separation techniques prior to reconstruction of the gravitational potential.  Our study of two such methods highlights the importance of broad frequency coverage for this purpose.  A companion paper presents application to the \Planck\ data on the ESZ-XMM sample.}


   \maketitle
 %

\section{Introduction}
The most important property of galaxy clusters for cosmological studies is their mass; it is also the most difficult to measure, because it is not directly observable.  Accurate mass measurements are needed, in particular, to calibrate scaling laws relating mass to observable cluster properties, such as richness \citep{yee2003,gladders2007,rozo2009}, X-ray properties \citep{arnaud2005,stanek2006, vikhlinin2006, arnaud2007, pratt2009, mantz2010, rozo2014b} or Sunyaev-Zeldovich (SZ) signal strength \citep{marrone2012, pip3_2013, bocquet2015, rozo2014c, WtGIV, hoekstra2015}.  The uncertainty in the mass calibration of these relations now limits the constraining power of cluster counts as a cosmological probe \citep{rozo2013, planck2013xx, hasselfield2013, reichardt2013, rozo2010, mantz2010, mantz2014,vikhlinin2009,bohringer2014,planck2014-a30}.  

Cluster masses can be determined dynamically, by application of the virial theorem to the velocity distribution of member galaxies, from X-ray studies of the intra-cluster medium, assuming hydrostatic equilibrium for the gas, and via the effects of gravitational lensing that distort the shapes of background galaxies.  Each approach presents its own advantages while suffering from specific systematic biases \citep{allen2011}.  

In this work, we discuss lensing of the cosmic microwave background (CMB) anisotropies as a promising new technique for measuring cluster masses, presenting a methodology for practical application with the specific aim of calibrating cluster scaling relations.  We evaluate the potential of current and future experiments to employ the methodology and show how to account for astrophysical biases from other cluster signals.  As illustration, we apply the technique to \Planck\ simulations of massive clusters and demonstrate that it is possible to recover an unbiased estimate of the mass scale of the \Planck\ XMM-Early Release SZ catalogue  \citep[ESZ-XMM,][]{planck2011viii, planck2011xi}.  A companion paper presents results obtained with the \Planck\ dataset for the ESZ-XMM sample.

Study of CMB lensing \citep[for a review see][]{lewis2006} is a rapidly growing field driven by the current generation of sensitive, high resolution CMB experiments.  Recent measurements of lensing in the CMB temperature power spectrum have been given by the Atacama Cosmology Telescope \citep[ACT,][]{das2014}, the South Pole Telescope \citep[SPT,][]{story2013} and the \Planck\ mission \citep{planck2013xvi}.  Direct reconstruction of the matter power spectrum using higher order statistics, designed to capture lensing's specific non-Gaussian mode coupling signature \citep{hu2002}, have also been reported by ACT \citep{das2011b} and SPT \citep{vanengelen2012}, as well as \Planck, which in addition produced an all-sky map of the projected matter distribution \citep{planck2013xvii}.

Lensing of the CMB by galaxy clusters was first discussed at length by \citet{zaldarriaga1999} when developing methods for reconstruction of the gravitational lensing potential.  \citet{seljak2000} considered instead the characteristic perturbation to the unlensed CMB temperature field, approximated locally as a pure gradient, induced by cluster lensing, an idea further studied by \citet{holder2004} and \citet{vale2004}.  In an approach analogous to the first authors, \citet{maturi2005} built a filter nonlinear in CMB temperature to reconstruct the lensing convergence field around clusters as a means of studying their density profiles.

In this paper, we develop a cluster mass extraction method based on a matched filter for the projected gravitational potential.  Our approach is similar in spirit to the work of \citet{zaldarriaga1999} and \citet{maturi2005}
in that we first reconstruct the lensing field around a cluster, rather than working with the lensing perturbation in the CMB temperature itself.  In the present work, however, we focus on determining total cluster mass even in relatively low signal-to-noise regimes.  Once we have the map of a cluster potential, obtained using the quadratic estimator of \citet{hu2002}, we then apply a matched filter designed to optimally extract the cluster's mass, assuming a density profile.  This allows us to obtain measurements for individual clusters, even when noise dominated, and use them in statistical analyses; for example, finding the mean mass of clusters by binning according to SZ signal strength or, in other words, the mass-SZ scaling relation.

In practice, we must confront a number of possible systematic biases.  We consider the effects of and ways to mitigate astrophysical signals contaminating the CMB map required as input for reconstruction of the gravitational potential.  The most difficult in this context are signals generated by the cluster itself, such as the thermal SZ~\cite[tSZ,][]{sz1972} and the kinetic SZ~\cite[kSZ,][]{sz1980} effects.  Component separation is therefore a crucial step prior to reconstruction of the potential, and our study will demonstrate the importance of multi-frequency observations in this context.  The kSZ effect, having the same spectral signature as the CMB, requires separate treatment; fortunately, we will see that the lensing potential reconstruction significantly reduces its impact by averaging it with uncorrelated CMB anisotropies.

We organize the paper as follows.  We begin in Sect.~\ref{sec:mass_estimation} by establishing our data model and discussing the reconstruction of the lensing potential and application of the matched filter; the presentation focuses at this point on the ideal case where we have a clean CMB map of known noise properties.  This allows a preliminary evaluation of the potential of current and future CMB experiments to measure cluster masses.  Section~\ref{sec:compsep} focuses on the tSZ and kSZ signals.  We employ two techniques to remove the tSZ signal and produce clean CMB maps from a set of individual frequency maps, and we evaluate the impact of the kSZ signal.  We simulate \Planck\ observations of a set of massive clusters in Sect.~\ref{sec:simulations} to illustrate the method, showing that it is possible (Sect.~\ref{sec:results}) to recover an unbiased estimate of the cluster mass scale for a sample similar to the \Planck\  ESZ-XMM catalogue \citep{planck2011xi}.  We finish with a discussion (Sect.~\ref{sec:discussion}) and our conclusions (Sect.~\ref{sec:conclusion}).
Throughout, we adopt a flat $\Lambda$CDM cosmological model with $H_0=70\,$km s$^{-1}$Mpc$^{-1}$ and $\Omega_{\rm M}=1-\Omega_\Lambda=0.3$. 

\section{Mass Estimation}
\label{sec:mass_estimation}
We first define our data model in the general context, and then present the matched filter mass estimation by focussing on the ideal case where we have a clean map of lensed CMB anisotropies with only instrumental noise.

\subsection{Cluster Signals}
Consider a patch of sky centered on a galaxy cluster of mass $\Mfive$ at redshift $z$.  We refer to cluster mass inside the radius $\Rfive$, interior to which the mean mass density is 500 times the critical density at the cluster's redshift, $\rho_{\rm c}(z)$; i.e., $\Mfive = 500 (4\pi/3)\rho_{\rm c}(z) \Rfive^3$.  The hot, gaseous intra-cluster medium (ICM) generates both tSZ and kSZ effects, and the cluster's (projected) gravitational potential, $\phi$, lenses the CMB anisotropies by bending light rays and displacing the apparent line-of-sight.  

We model the mass distribution with a NFW \citep{nfw1996} profile,
\begin{equation}
\rho(r) = \frac{\rhoo}{(r/\rs)(1+r/\rs)^2},
\end{equation}
described by its central density, $\rhoo$, and physical scale $\rs$.  The latter can be related to $\Rfive$ using the concentration parameter, $\cfive$, as $\rs =\Rfive/\cfive$.  Unless otherwise stated, we take $\cfive=3$ in this work; in reality, it is expected to depend weakly on cluster mass and redshift, i.e., $\cfive(\Mfive, z)$~\citep{bullock2001,neto2007,munoz2011}.  Integrating along the line-of-sight yields the projected surface mass density at angular position $\vec{x}$ from the center,
\begin{equation}\label{eq:masstemplate}
\Sigma(\vec{x}) = \phio \; \Sigmam(x/\thetam), 
\end{equation}
where $x=|\vec{x}|$ and $\Sigmam$ is a template characterized by the angular scale $\thetam=\rs/\Dclus$, with $\Dclus$ the angular diameter distance to the cluster. The integral along the line-of-sight is performed out to $r=5\Rfive$. The normalization, $\phio$, is given below.

For the pressure of the ICM, we employ the modified NFW profile of \citet{nagai2007} with parameters given by the X-ray observations of \citet{arnaud2010}, the so-called universal pressure profile in the non self-similar case (Eq.~12 of that paper).   Integrating the pressure profile along the line-of-sight gives the tSZ angular template, $\Tth(x/\thetat)$, characterized by its scale radius, $\thetat$, and which is normalized by the cluster's central Compton $y$ value, $\yo$, to obtain the complete tSZ profile.  The kSZ signal is proportional to the optical depth through the cluster with profile $\Tk(x/\thetak)$ characterized by the scale radius $\thetak$.  It is normalized by the cluster peculiar velocity, $\beta$, to obtain the kSZ signal.  In our numerical calculations below, we use the same radial profile for the ICM pressure and optical depth, which is equivalent to approximating the gas as isothermal; this is not strictly the case, but the approximation has little effect on our conclusions.

We suppose that the region is observed in several millimeter/sub-millimeter bands, producing a set of maps at $N$ different frequencies $\nui$ ($i=1,...,N$) that we arrange in a column vector $\vec{m}(\vec{x})$, a function of angular position $\vec{x}$ on the sky and whose $i^{th}$ component is the the map $m_i(\vec{x})$ at frequency $\nui$.  We assume the maps are in units of thermodynamic temperature, so the CMB and kSZ signals remain constant across frequencies, and we denote the beam at frequency $\nu_i$ by $b_i$.  

The maps contain the cluster tSZ and kSZ signals, lensed CMB anisotropies and noise, 
\begin{equation}
\label{eq:datamodel}
\vec{m}(\vec{x}) =  \yo\vec{\tth}(\vec{x}) + \beta\vec{\tk}(\vec{x}) + \vec{s}(\vec{x}) + \vec{n}(\vec{x}),
\end{equation}
where $\vec{\tth}(\vec{x})$ is the vector whose $i^{th}$ component is \mbox{$\jnu(\nu_i) [b_i\ast \Tth](\vec{x})$}, the beam-convolved tSZ template modulated by the tSZ frequency spectrum, $\jnu$, in temperature units.  The components of the kSZ vector, $\vec{\tk}(\vec{x})$, are \mbox{$[b_i\ast \Tk](\vec{x})$}, and those of the CMB vector, $\vec{s}(\vec{x})$, are $s_i(\vec{x})= [b_i\ast S](\vec{x})$, where we denote the CMB signal on the sky as $S(\vec{x})$. As for $\Sigmam$, the integration along the line-of-sight for $\Tth$ and $\Tk$ is performed out to $r=5\Rfive$.

The unlensed (and unobservable) CMB field, ${\tilde S}(\vec{x})$, is transformed into the observed sky signal as \citep[e.g.,][]{bartelmann2001}
\begin{eqnarray}
S(\vec{x}) & = & {\tilde S}(\vec{x}) + \delta S(\vec{x}),\\
\delta S(\vec{x}) & = & \nabla S(\vec{x}) \cdot \nabla \phi(\vec{x}),
\end{eqnarray}
to first order in the lensing potential, which is related to the convergence,
\begin{equation}
\kappa(\vec{x})= \Sigma({\vec{x}})/\Sigmacr,
\end{equation}
by 
\begin{eqnarray}
\phi(\vec{x}) & = &\int d^2x' \; \kappa(\vec{x'})  \ln(|\vec{x}-\vec{x}'|) \\
                      & = & \phio \int d^2x'\; \frac{\Sigmam(x'/\thetam)}{\Sigmacr} \ln \left|\vec{x}-\vec{x}'\right| \\
\label{eq:lenspotential}
                      & \equiv &\phio\Phi(x/\thetam).
\end{eqnarray}
All integrals are restricted to $x'<5\theta_{500}$.  The critical surface mass density, $\Sigmacr={c^2 \over 4 \pi G} {\DCMB \over \Dclus D_{\rm clus - CMB}}$, is defined in terms of the angular diameter distances $\DCMB$, $\Dclus$ and $D_{\rm clus - CMB}$ between the observer and the CMB, the observer and the cluster, and the cluster and the CMB, respectively.  The third equality in Eq.~(\ref{eq:lenspotential}) defines our (dimensionless) model template for the lensing potential,  $\Phim$, parameterized by the angular scale $\thetam$.  It is normalized by $\phio$ (see Eq.~\ref{eq:masstemplate}), defined such that $\Phim(0)=1$.

The noise term in Eq.~(\ref{eq:datamodel}), $\vec{n}(\vec{x})$, includes instrumental noise and astrophysical signals that are not related to the cluster.  Examples of the latter are Galactic foreground emission, extragalactic point sources and lensing by matter randomly projected along the line-of-sight (large-scale structure, or LSS, noise).  The effects of LSS correlated with cluster position can only be evaluated with numerical simulations and remain beyond the scope of the present work.  Similarly, the background extragalactic point source population is modified near the cluster by lensing, creating a second order cluster-related signal that we do not consider in this work.  

\subsection{Reconstruction of the Lensing Potential Map}
Given a map of CMB temperature anisotropy, $\hat{S}$ -- obtained from a prior component separation step, as describe below -- centered on a cluster, we apply the flat-sky quadratic estimator from \citet{hu2002} to find the Fourier modes of the projected gravitational potential:
\begin{equation}
\label{eq:huok}
\estphi(\vec{K}) = A(\vec{K}) \sum_{\vec{k}} \hat{S}^*(\vec{k}) \hat{S}(\vec{k'}) F(\vec{k},\vec{k'}),
\end{equation}
with $\vec{K}=\vec{k}-\vec{k}' {\rm [mod \, n]}$, where n is the number of pixels along the x (or y) axis, and
\begin{equation}
\label{huok_err}
A(\vec{K}) = \left [ \sum_{\vec{k}} f(\vec{k},\vec{k}') F(\vec{k},\vec{k}') \right ]^{-1}.
\end{equation}
The weights $F(\vec{k},\vec{k}')$ are defined so that $\estphi$ is the minimum variance estimator:
\begin{equation}
F(\vec{k},\vec{k}') = {f^*(\vec{k},\vec{k}') \over 2 P_{\hat{S}}(k) P_{\hat{S}}(k')},
\end{equation}
in which $P_{\hat{S}}(k) = |b(k)|^2 C_k +  P_{\rm noise}(k)$ is the observed power spectrum with a contribution from the effective noise, $P_{\rm noise}$, of the cleaned CMB map; $b(k)$ represents the effective instrumental beam\footnote{Taken, for simplicity, to be axially symmetric.}; $C_k$ is the power spectrum of the (true) lensed CMB signal, $S(\vec{x})$; and $f(\vec{k},\vec{k})$ is given by
\begin{equation}
f(\vec{k},\vec{k}') = b^*(\vec{k}) b(\vec{k}') \left [ {\tilde C}_{k} \vec{k} \cdot \vec{K} - {\tilde C}_{k'}  \vec{k}' \cdot \vec{K} \right ],
\end{equation}
where ${\tilde C}_{k}$ is the power spectrum of the unlensed CMB sky, ${\tilde S}({\vec x})$.

Equation~(\ref{eq:huok}) gives us an unbiased estimate of $\phi(\vec{K})$ for a cluster of given properties in the sense that if averaged over all realizations of the (unlensed) CMB and instrumental noise, $\langle \estphi(\vec{K}) \rangle = \phi(\vec{K})$.  The variance of the estimate about the mean is given by $A(\vec{K})$, and the reconstruction is optimal in that it minimizes this variance for each mode $\vec{K}$.  

\subsection{Matched Filter}
Adopting the model potential template of Eq.~(\ref{eq:lenspotential}), our matched filter operates on the lensing potential map to extract the normalization, $\phio$, for a given scale $\thetam$.  Each mode $\vec{K}$ of the estimated potential, $\estphi$, is an independently measured variable with standard deviation $A^{1/2}(K)$.  We therefore construct the matched filter for the potential amplitude as
\begin{equation}
\label{eq:estphio}
\estphio = \left [ \sum_{\vec{K}} {| \Phi(\vec{K}) |^2 \over A(\vec{K})} \right ]^{-1} \sum_{\vec{K}} {\Phi^*(\vec{K}) \over A(\vec{K})}  \estphi(\vec{K}),
\end{equation}
where $\Phi(\vec{K})$ is the Fourier transform of the model template (Eq. ~\ref{eq:lenspotential}).
This yields an unbiased estimate of $\phio$ with minimal variance given by
\begin{equation}
\label{eq:varphio}
{\rm Var}(\estphio) = \left [ \sum_{\vec{K}} {| \Phi(\vec{K}) |^2 \over A(\vec{K})} \right ]^{-1}.
\end{equation}

Once normalized by our measurement, $\estphio$, the cluster mass model is completely specified.  We could quote our filter measurements directly as $\estphio$, but choose instead to express them in terms of the integrated convergence calculated using the model:
\begin{equation}
\label{eq:Kappafiver}
\Kappafiver  \equiv  2\pi \int_0^{5\theta_{500}} dx \; x \kappa(x) = \frac{1}{\Dclus^2(z) \Sigmacr(z)} \Mfiver.
\end{equation}
Note that $\theta_{500}=R_{500}/\Dclus(z)=\cfive\thetam$.  The first equality defines our preferred observable, and the second relates it to cluster mass calculated within the radius $5\Rfive$; this is easily translated into $\Mfive$ given the concentration parameter, $\cfive$.  Our estimator for this observable is
\begin{eqnarray}
\label{eq:estKappafiver}
\estKappafiver & \equiv &2\pi \estphio \int _0^{5\theta_{500}} dx \; x \frac{\Sigmam(x/\thetam)}{\Sigmacr(z)}\\
   & = & \frac{1}{\Dclus^2(z) \Sigmacr(z)} \estMfiver
\end{eqnarray}
Hereafter, we consider $\estKappafiver$ as the output of the filter; it is dimensionless and expressed in arcmin$^2$.  The second line defines our cluster mass estimator, whose units are set by the prefactor.

These estimators are unbiased over the CMB and noise ensembles: $\langle\estKappafiver\rangle = \Kappafiver$ and $\langle\estMfiver\rangle=\Mfiver$.  They are also optimal in that they minimize their respective variances over the same ensemble.  Explicitly, we have
\begin{equation}
\label{eq:varMfiver}
{\rm Var}(\estMfiver) = \left[\Dclus^2(z)\Sigmacr(z)\right]^2\sigmaKappa^2,
\end{equation}
for the variance of the mass estimator, where $\sigmaKappa^2$ is the variance of our filter output, calculated using Eqs.~(\ref{eq:varphio}) and (\ref{eq:estKappafiver}).  This is the uncertainty on the mass measurement of a single cluster, with contributions from instrumental noise and the CMB fluctuations themselves.  We will see below that the kSZ adds an additional source of noise, as well as a bias term; fortunately, they are small in practice.

\begin{figure}
\centering
\includegraphics[width=.5\textwidth]{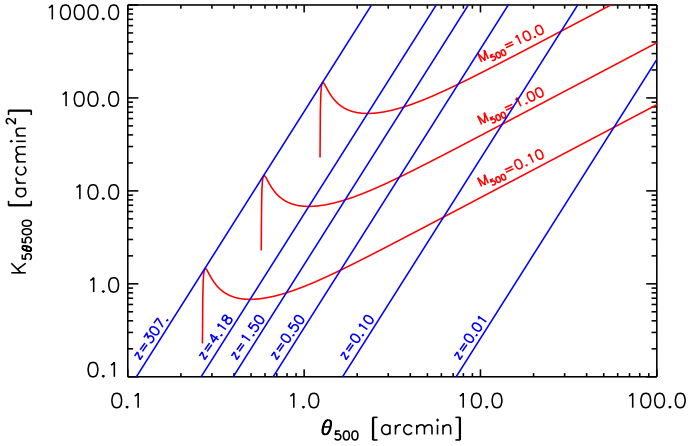}
\caption{Contours of cluster mass in $10^{15} \Msolar$ units (red curves) and redshift (blue curves) projected onto the observational plane defined by the filter output $\Kappafiver$ and angular scale $\theta_{500}=\cfive\thetam$.  A cluster of fixed mass $\Mfive$ follows a red contour according to Eq.~(\ref{eq:Kappafiver}) as it moves out in redshift.  
Each mass contour follows the same pattern, simply displaced in amplitude.  This figure shows how each point in the observational plane maps to a point in the cluster plane of $(\Mfive,z)$.}
\label{fig:isoMz}
\end{figure}

Figure~\ref{fig:isoMz} illustrates the relation between the observation plane $(\Kappafiver, \theta_{500})$ and physical cluster quantities.  Each point in this plane maps directly to a point in the cluster $(\Mfive,z)$ plane, as shown by the contours of iso-mass (in red) and iso-redshift (blue).  At fixed mass, Eq.~(\ref{eq:Kappafiver}) specifies the evolution of $\Kappafiver$ as a function of redshift, which when coupled with $\theta_{500}\propto \Mfive^{1/3}/\Dclus(z)$ describes an iso-mass curve.  As the cluster moves out in redshift, its angular size decreases; at the same time, the prefactor in Eq.~(\ref{eq:Kappafiver}) determines the decrease and final upturn in the integrated convergence.  The rapid decrease from low redshift outwards may seem surprising, but is due to the fact that we integrate over angular extent.  Integrated over physical extent, the convergence increases at first with redshift as the lensing kernel becomes more efficient, but the angular extent decreases and drives down the value of $\Kappafiver$.  Each mass follows the same general curve, simply displaced in absolute scale.  

\begin{figure}
\centering
\includegraphics[width=.5\textwidth]{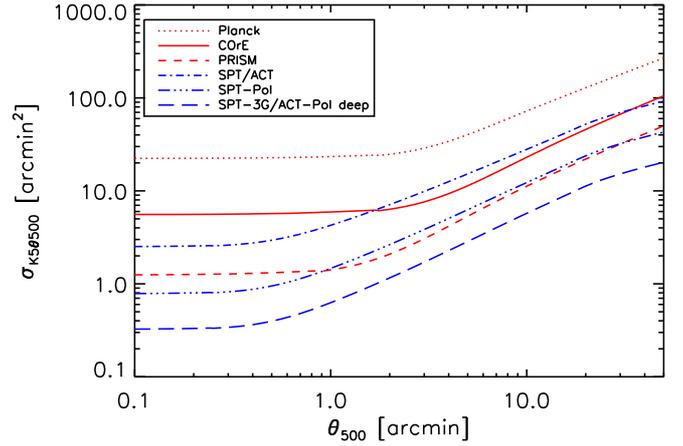}
\caption{Standard deviation, or filter noise, of the matched filter output (Eq.~\ref{eq:Kappafiver}) versus filter scale $\theta_{500}=\cfive\thetam$ for the different experimental setups, as labeled.  The red curves give results for the \Planck\ satellite and two future missions proposed to ESA, COrE and PRISM.  The first has similar angular resolution to \Planck\ ($\sim 5$\, arcmin FWHM), but lower noise, while PRISM has both lower noise and higher resolution at $\sim 2.6$\, arcmin.  The blue curves show the noise levels for SPT/ACT, SPT-Pol and SPT-3G/ACT-Pol.  All curves assume the filter is perfectly matched to the clusters.  The experimental characteristics are summarized in Table~\ref{tab:expts}.}
\label{fig:obsplane}
\end{figure}

A given experiment will trace a sensitivity curve in the observation plane.  In Figure~\ref{fig:obsplane} we show predictions for the filter sensitivity, expressed by the standard deviation of the filter variance, as a function of filter angular scale for a number of different experiments.  The top three curves (traced in red) all refer to space-based experiments, while the three lower curves (in blue) represent ground-based experiments similar to the three generations of SPT~\citep{story2013,austermann2012,benson2013}, ACT~\citep{das2014} and ACT-Pol~\citep{niemack2010}.  The space missions are \Planck~\citep{planck2013i} and two missions proposed to the European Space Agency, COrE~\citep{core2011} and PRISM~\citep{prism2013}.  The former has similar resolution to \Planck, but more detectors and lower noise, while the latter envisions a larger telescope with both lower noise and higher angular resolution.  

We summarize the adopted characteristics of each experiment in Table ~\ref{tab:expts} in terms of angular resolution and white noise in the reconstructed CMB map\footnote{In each case, the reconstructed CMB map was assigned the characteristics of the experimental band closest to 143~GHz.}.  We emphasize, however, that the experiments have very different frequency coverage.  Although not accounted for in this present discussion, we show later that extensive frequency coverage is crucial for proper CMB reconstruction and, especially, removal of the tSZ signal.

All sensitivity curves start on small angular scales with a flat response and then break to a rise toward larger filter scales.  The break occurs on smaller scales for the higher resolution ground-based experiments.  COrE and \Planck\ have the same resolution and break on the same scale, but with its lower noise level, COrE's plateau lies below that of \Planck.  PRISM has a noise level comparable to SPT-3G, but slightly lower angular resolution.  We see that it breaks at an intermediate filter scale and on a higher plateau than SPT-3G.  This demonstrates the interplay of angular resolution and noise:  At a given noise level in $\mu$K-arcmin, an experiment with higher angular resolution accesses more modes $\vec{k}$ to reconstruct a given potential mode $\vec{K}$ (Eq.~\ref{eq:huok}), thereby reducing the filter noise.  

\begin{table*}[thb] 
  \begingroup
  \newdimen\tblskip \tblskip=5pt
  \caption{Characteristics of representative surveys: \Planck\ \citep{planck2013i}, COrE \citep{core2011}, PRISM \citep{prism2013}, ACT \citep{das2014}, ACT-Pol \citep{niemack2010}, SPT \citep{story2013}, SPT-Pol \citep{austermann2012}, SPT-3G \citep{benson2013}}
  \label{tab:expts}
  \vskip -3mm
  \footnotesize
  \setbox\tablebox=\vbox{
  \newdimen\digitwidth
  \setbox0=\hbox{\rm 0}
  \digitwidth\wd0
  \catcode`*=\active
  \def*{\kern\digitwidth}
  \newdimen\signwidth
  \setbox0=\hbox{+}
  \signwidth=\wd0
  \catcode`!=\active
  \def!{\kern\signwidth}
  \halign{#\hfil\tabskip=2em& #\hfil&\hfil #\hfil& \hfil#\hfil\tabskip=0pt\cr
  \noalign{\doubleline}
  \omit\hfil Name\hfil&\omit\hfil Location\hfil&\omit\hfil Map Resolution (FWHM) \hfil&\omit\hfil Map Noise\hfil\cr
  \omit\hfil \hfil&\omit\hfil \hfil&\omit\hfil [arcmin] \hfil&\omit\hfil [$\mu$K-arcmin]\hfil\cr
      \noalign{\vskip 3pt\hrule\vskip 5pt}
      \Planck\dotfill& Space& 5.0& 45.0\cr
      COrE\dotfill& Space& 5.0&    *2.6\cr
      PRISM\dotfill& Space& 2.5&  *2.6\cr
      Generation: SPT/ACT\dotfill& Ground& 1.0& 18.0\cr
      Generation: SPT-Pol\dotfill& Ground& 1.0& *5.0\cr
      Generation: SPT-3G/ACT-Pol deep\dotfill& Ground& 1.0& *2.0\cr
  \noalign{\vskip 5pt\hrule\vskip 3pt}}}
  \endPlancktablewide 
  \endgroup
\end{table*}

\begin{figure}
\centering
\includegraphics[width=.5\textwidth]{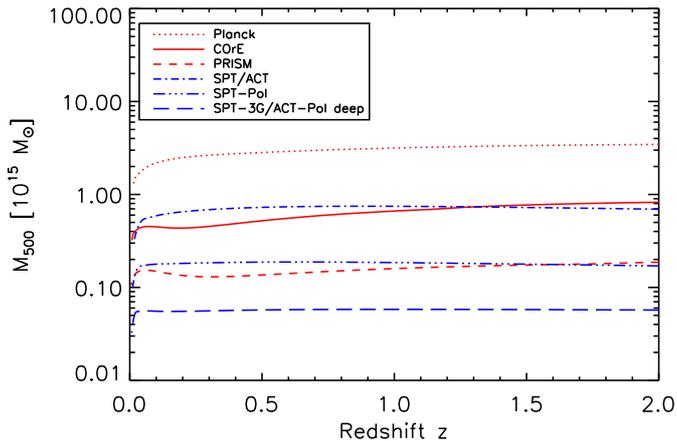}
\caption{Standard deviation of mass measurements, $\Mfive(z)$, with the matched filter (square root of Eq.~\ref{eq:varMfiver}) as a function of cluster redshift for the same experimental setups plotted in Fig.~\ref{fig:obsplane}.  This comparison does not take into account the ability of each experiment to eliminate contaminating signals, such as the tSZ, which depends on spectral coverage.  In this context, the space-based missions benefit from wider spectral coverage (see text).}
\label{fig:Mlim}
\end{figure}

In Fig.~\ref{fig:Mlim} we give the standard deviation of the mass measurement, $\Mfive(z)=\sigma(z)$ (square root of Eq.~\ref{eq:varMfiver}), as a function of redshift for each experimental setup.  The results provide a useful metric for each experiment's ability to measure cluster mass: The uncertainty on the mean mass for a sample of $N$ clusters at redshift $z$ will be $\Mfive(z)/N^{1/2}$.  Note, however, that a cluster of $n\Mfive(z)$ will have a significance smaller than $n\sigma$.  This is because as the mass increases from the limiting value, the cluster's angular size also increases, driving the filter noise higher (unless we are on the small-scale plateau of the curves in Fig.~\ref{fig:obsplane}).

We see that COrE and SPT/ACT can only be expected to measure mass for the most massive systems in the universe, while \Planck\ cannot be expected to measure any individual cluster mass.  The sensitivity of these experiments, however, is sufficient to obtain mean mass as a function of other cluster observables by binning measurements; in other words, to establish mean observable-mass scaling relations.  Herein lies the value of our matched filter approach, by providing a means of combining many low signal-to-noise measurements to statistically determine cluster mass.  The sensitivity of SPT-Pol, SPT-3G/ACT-Pol and the PRISM mission, on the other hand, is sufficient to enable individual cluster mass measurements as well as statistical studies.

\section{Astrophysical Contaminants} 
The above discussion supposes that we have a clean map of CMB anisotropies from which to extract the lensing signal.  To produce this map, we must first separate the CMB component from other astrophysical emission.  In this work, we focus on the potentially most troublesome signals, those produced by the cluster itself at CMB-dominant frequencies, namely the tSZ and kSZ effects.

Generally, we employ the Internal Linear Combination (ILC) methodology to separate the CMB from other signals.  Because we will find that  the standard ILC does not sufficiently remove the tSZ signal, causing an important bias in our mass estimation, we present two techniques based on the ILC that seek in addition to actively remove the tSZ signal: template fitting and subtraction of the tSZ, and an ILC constrained to cancel the tSZ.  Both prove satisfactory.

The kSZ cannot be eliminated in this fashion, however, having the same spectral signature as the CMB.  It could be removed by template fitting for experiments with sufficient resolution, but as will be seen, it does not produce any significant bias in our mass estimations.  Keeping this in mind, we restrict ourselves to an examination of the nature of the kSZ contamination on the reconstructed lensing potential.  This will guide our interpretation of the simulation results.  

We test our methodology using the simulations described in Sect.~\ref{sec:simulations}, presenting the results in Sect.~\ref{sec:results}.  We will see the importance of actively removing the tSZ, and that any effect from the kSZ remains manageable.  Our simulations do not include diffuse Galactic or extragalactic emission not related to the clusters.  At this point of our study, having employed an ILC at the heart of our component separation techniques, we assume that these are adequately controlled; the issue will be studied in more detail in future work.

\label{sec:compsep}
\subsection{The Thermal SZ Signal}
We develop two approaches to removing the tSZ signal:  multifrequency matched filters (MMF) to estimate the amplitude of the tSZ effect and then remove it from the survey frequency maps prior to a standard ILC, and an ILC constrained to eliminate the tSZ signal while extracting the CMB map.  We compare the performance of the two approaches in Sect.~\ref{sec:results}.  The goal is to produce the CMB map, $\hat{S}$, used as input in the lensing reconstruction of Eq.~(\ref{eq:huok}).

\subsubsection{Multifrequency Matched Filters (MMF)}
With this approach we apply the MMF of \citet{melin2006} using the pressure profile $\Tth(x/\thetat)$  to obtain an estimate of the central Compton $y$ parameter:
\begin{equation}
\estyo = \int d^2x \; \vec{\Psit}^t(\vec{x}) \cdot \vec{m}(\vec{x}),
\end{equation}
where
\begin{equation}
\vec{\Psit}(\vec{k}) = \sigt^2 \vec{P}^{-1}(\vec{k})\cdot \vec{\tth}(\vec{k}),
\end{equation}
with
\begin{eqnarray}
\label{eq:sigt}
\sigt          & \equiv & \left[\int d^2k\; 
     \vec{\tth}^t(\vec{k})\cdot \vec{P}^{-1} \cdot
     \vec{\tth}(\vec{k})\right]^{-1/2},
\end{eqnarray}
$\vec{P}(\vec{k})$ being the interband power spectrum matrix with contributions from (non-tSZ) sky signal and instrumental noise.  It is the effective noise matrix for the MMF and can be estimated directly on the data, since the tSZ is small compared to other astrophysical signals in the sky patch centered on the cluster. 

We assume that the tSZ is accurately described by our model and seek only to normalize the tSZ template to each cluster.  This is the case for our simulations, because we are using the same profile for the simulated tSZ and the filter template.  In reality, possible mismatch between the true cluster profile and the filter template would be a source of uncertainty. 

Once we have normalized the template, we remove the tSZ signal from each of the $N$ individual frequency maps,
\begin{equation}
\vec{\widehat{m}}(\vec{x}) =  \vec{m}(\vec{x}) - \estyo \vec{\tth}(\vec{x}),
\end{equation}
and apply a standard ILC to reconstruct the clean CMB map:
\begin{equation}
\hat{S}(\vec{k})=  \left [ \vec{b}^t \vec{P}^{-1} \vec{b} \right ]^{-1}  \vec{b}^t \vec{P}^{-1} \vec{\widehat{m}}(\vec{k}),
\end{equation}
where $\vec{b}$ is the beam vector of dimension $N$.  Finally, we convolve the resulting map with a fiducial beam, $b_{\rm fid}$ (5\,arcmin in the case of Planck), to obtain the $\hat{S}(\vec{k})$ used in Eq.~(\ref{eq:huok}).  The power spectrum associated with this map is
\begin{equation}
P_{\hat{S}}= |b_{\rm fid}|^2 \left [ \vec{b}^t \vec{P}^{-1} \vec{b} \right ]^{-1}.
\end{equation}

\subsubsection{Constrained Internal Linear Combination (CILC)}
In the second approach, we directly construct the clean CMB map with a constrained ILC designed to nullify the tSZ effect, making use of its well-defined spectral signature, $\jnu$.  The formalism is described in detail by~\citet{remazeilles2011}.  The reconstructed clean CMB map can be written
\begin{eqnarray}
\hat{S}(\vec{k})=\Delta^{-1} \left[ \left ( \vec{\jnu}^t \vec{P}^{-1} \vec{\jnu} \right ) \vec{b}^t  \vec{P}^{-1} -  \left ( \vec{b}^t \vec{P}^{-1} \vec{\jnu} \right ) \vec{\jnu}^t  \vec{P}^{-1} \right] \vec{m}(\vec{k}), \\
\Delta = \left ( \vec{b}^t \vec{P}^{-1} \vec{b} \right ) \left ( \vec{\jnu}^t \vec{P}^{-1} \vec{\jnu} \right ) - \left ( \vec{b}^t \vec{P}^{-1} \vec{\jnu} \right )^2,
\end{eqnarray}
with $(\vec{\jnu})_i\equiv \jnu(\nui) b_i$.  The noise matrix, $\vec{P}$, is constructed as before.

We again convolve the resulting map with the fiducial beam, $b_{\rm fid}$, to obtain the CMB map, $\hat{S}(\vec{k})$, used in Eq.~(\ref{eq:huok}).  The power spectrum for this fiducial map is
\begin{eqnarray}
P_{\hat{S}} = \frac{|b_{\rm fid}|^2}{\Delta^{2}}
\left[  \left ( \vec{\jnu}^t \vec{P}^{-1} \vec{\jnu} \right )^2 \left ( \vec{b}^t \vec{P}^{-1} \vec{b} \right ) - \left ( \vec{\jnu}^t \vec{P}^{-1} \vec{\jnu} \right )  \left ( \vec{b}^t \vec{P}^{-1} \vec{\jnu} \right )^2  \right].
\end{eqnarray}

The strength of the CILC removal, compared to the MMF approach, is its insensitivity to cluster modeling uncertainties (e.g., SZ profile).  On the other hand, the variance of the noise in the reconstructed CMB map is higher, slightly increasing the error on the final mass estimate, as we show in Sec.~\ref{sec:results}. We also present the impact of tSZ modeling uncertainties on the MMF approach in the same section.

\subsection{The Kinetic SZ Signal}
\label{sec:kSZ}
After the tSZ, the kSZ is the dominant cluster signal at CMB frequencies.  It has the same spectrum as the primary CMB and cannot be removed by spectral separation methods.  

We can appreciate the effect of the kSZ by returning to the lensing potential reconstruction of Eq.~(\ref{eq:huok}).  Even in the best possible case, the reconstructed CMB map, $\hat{S}$, contains kSZ in addition to the lensed CMB and noise: $\hat{S}(\vec{x})=S(\vec{x})+\beta\Tk(x/\thetak)+n(\vec{x})$.  The lensing map reconstruction therefore has four contributions:
\begin{eqnarray}
\estphi'(\vec{K}) & = & \estphi(\vec{K}) + 2{\cal R}\left[\beta A(\vec{K})\sum_{\vec{k}}S(\vec{k})\Tk^*(\vec{k}'\thetak)F(\vec{k},\vec{k}')\right]  \\
 & & + 2{\cal R}\left[\beta A(\vec{K})\sum_{\vec{k}}n(\vec{k})\Tk^*(\vec{k}'\thetak)F(\vec{k},\vec{k}')\right] \\
 & & + \beta^2 A(\vec{K})\sum_{\vec{k}}\Tk(\vec{k}\thetak)\Tk^*(\vec{k}'\thetak)F(\vec{k},\vec{k}')\\ 
 \label{eq:estphikSZ}
 & \equiv & \estphi(\vec{K}) + \beta\left[\estphiks(\vec{K}) + \estphikn(\vec{K})\right] + \beta^2\estphikk(\vec{K})
\end{eqnarray}
where $\estphi$ contains just the CMB and noise terms, as considered in Sect.~\ref{sec:mass_estimation}.  The kSZ adds cross terms of the kSZ with both CMB, $\estphiks$, and noise, $\estphikn$, and a term quadratic in the kSZ signal, $\estphikk$.  

Averaged over the CMB and noise ensembles, the two middle terms vanish, independent of the value of $\beta$, because $\langle s(\vec{k})\rangle = \langle n(\vec{k})\rangle=0$.  They act as an additional noise contribution to the potential reconstruction for a given cluster.  This behaviour differs from that of the kSZ when directly using the temperature anisotropy induced by cluster lensing, rather than reconstruction of the lensing potential as done here.  For a given $\beta$, the kSZ is guaranteed to contribute to the temperature anisotropy at a level comparable to the lensing signal \citep{seljak2000,lewis2006}; in our case, however, the contribution could be small, depending only on the chance alignment of CMB and kSZ modes.  In either case, additional averaging over a set of clusters will further reduce the effect of these terms linear in $\beta$ because the objects will have random peculiar velocities.  

The last term, quadratic in $\beta$, is a bias.  Its presence is independent of the CMB and noise ensembles, and it cannot be beaten down by averaging over a cluster ensemble.  With proper modeling of the kSZ signal, the bias could be eliminated, if needed, through subtraction, cluster by cluster, to leave a zero-mean residual as a noise contribution.  This will not prove necessary in our subsequent study, where we will find that the bias is unimportant for realistic cluster velocities.  

\section{Simulations}
\label{sec:simulations}
We illustrate our mass estimation technique through recovery of the mass scale for a sample of clusters with simulated \Planck-like  observations. We proceed by first simulating a sample of identical clusters, and then consider a mock of the ESZ-XMM, a subsample of 62 clusters from the \Planck\ Early Sunyaev-Zeldovich list \citep[ESZ,][]{planck2011viii} with good X-ray observations, including X-ray determined masses, $\Mx$, spanning the range  $[2-20]\times 10^{14}\, \Msolar$ \citep[ESZ-XMM,][]{planck2011xi}.  In a companion paper, we report an estimation of the mass scale of the actual ESZ-XMM sample using the \Planck\  dataset.

Our first simulation consists of 62 observations of a mock of A2163, assigning mass and tSZ profiles following our adopted templates.  With an X-ray deduced mass of $\Mx=1.9\times 10^{15}\, \Msolar$ ($z=0.203$), this system is one of the most massive clusters known, falling near the $1\sigma$ curve for \Planck\ shown in Fig.~\ref{fig:Mlim}, and the most massive member of the ESZ-XMM.  

We generate 62 independent realizations of primary CMB anisotropies and of white noise in tangential map projections of  $10 \times 10 \deg^2$ across the six highest frequency channels (100-857\, GHz) of \Planck.  The mock of A2163 is centered in each channel map, the lensing is applied to the primary CMB anisotropies and the tSZ signal added.  We then smooth each channel map by its corresponding beam and add the white noise, taking the noise and beam characteristics as published in \citet{planck2013i}.  A simulated cluster observation thus comprises six channel maps, and there are 62 such simulated observations.

For each of these mock observations, we first remove the tSZ signal and produce a clean CMB map as described above.  We then reconstruct the lensing potential map and apply the matched filter to extract our mass estimate, $\estMfive$.  Note that the filter is perfectly matched to the cluster in that the filter template and actual cluster projected potential are identical.  We refer to the complete processing of a single cluster observation as an analysis {\em chain}. 
 
For each chain, we compare the mass measurement to the input (X-ray deduced) mass of A2163 by forming the ratio $\estMfive/\Mx$, and then take the sample mean, $\langle \estMfive/\Mx\rangle$, over the 62 observations.  An unbiased recovery of the sample mass scale corresponds to $\langle \estMfive/\Mx\rangle=1$.  

We run two sets of 62 chains without kSZ to compare the results from the two different component separation methods presented in Sect.~\ref{sec:compsep}.  We also ran additional simulation chains adding the kSZ effect with constant (systematic) peculiar velocities of 300 \mbox{km\,s$^{-1}$} and 900 \mbox{km\,s$^{-1}$} to each of the 62 clusters.

\section{Results}
\label{sec:results}
\begin{figure}[htb]
\begin{center}
\includegraphics[width=0.5\textwidth]{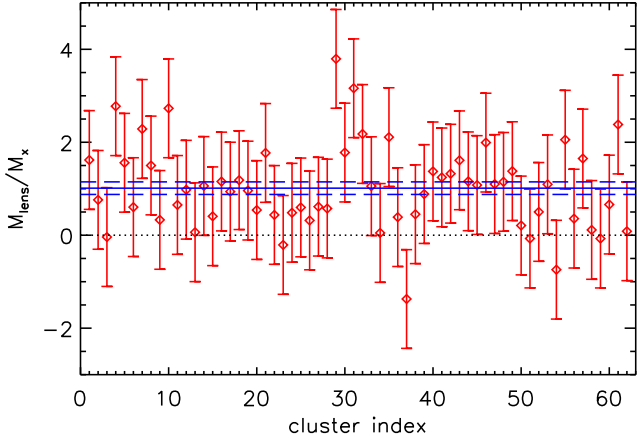}
\end{center}
\caption{Recovered mass for the 62 \Planck\ simulations of A2163, expressed as the ratio of the CMB-measured mass to the input mass from the X-ray model, $\estMfive/\Mx$.  The tSZ was removed in this example using the CILC.  Each diamond is the result of a single analysis chain (simulation, CILC, lensing extraction and matched filtering) accompanied by its $1\sigma$ uncertainty of 1.06. The solid blue line shows the sample average (1.01) and its $1\sigma$ range ($\pm 1.06/\sqrt{62}=\pm 0.13$).}
\label{fig:result_2dilc}
\end{figure}

Figure~\ref{fig:result_2dilc} shows the recovered mass ratio $\estMfive/\Mx$ for the 62 simulated cluster observations without kSZ when using the constrained ILC to remove the tSZ.  The individual measurement uncertainty on this quantity is large at 1.06 (the same for all chains, because the cluster and the statistical CMB and noise properties are the same for the 62 realizations).  The points clearly disperse preferentially above zero, and taking the sample average we find $\langle \estMfive/\Mx\rangle = 1.01\pm0.13$. 

We have an unbiased recovery of the sample mass scale with 13\% uncertainty.  The result is identical when using the MMF component separation procedure\footnote{The individual measurement uncertainty is slightly lower in this case (1.04 against 1.06), but the sample mean and its uncertainty are the same to the given precision.}.  The basic result of this study, therefore, is that we detect the mass scale of the sample at greater than $7\sigma$.

If we model the SZ emission with the cool-core profile given in Table C.2. of~\cite{arnaud2010} but extract it with the MMF based on the universal profile, the sample average increases from $1.01\pm0.13$ to $1.19\pm0.13$. Adopting instead the morphologically disturbed profile, the value shifts to $1.09\pm0.13$. The mis-modeling of the SZ profile in the MMF component separation procedure may thus introduce a bias of order $\sim 1 \sigma$ on the average.

We have also tested the sensitivity of our conclusions to the extension of the mass profile by truncating the integration at $R_{vir}$ instead of $5\Rfive$.  The individual measurement uncertainty increases to 1.23, which in turn slightly increases the uncertainty on the sample average to 0.16.  This results in a modest decrease in the global significance of the detection from $7\sigma$ to between $6$ and $7\sigma$.

Accurate removal of the tSZ signal is essential, something which can be gauged from the results when applying a standard ILC to extract the CMB without any constraint to nullify the tSZ.  In this case, we find a sample mean of $\langle \estMfive/\Mx\rangle =1.93\pm0.13$, highly biased by the residual tSZ signal.  

The standard ILC is incapable of removing the tSZ signal to a sufficiently high level.  We expect that this is in large part due to the fact that the tSZ is only a weak component in the map and hence not accounted for by the standard ILC weights.  Our component separation techniques manage to adequately remove the tSZ by direct subtraction (MMF) or cancelation (CILC), both relying on the known spectral dependence of the signal.  It is clear that multi-band CMB observations for accurate removal of the tSZ are an important consideration in designing experimental campaigns.  

Turning to the kSZ effect, we find a sample mean of $\langle \estMfive/\Mx\rangle = 1.00\pm 0.13$ for the case of 300 \mbox{km\,s$^{-1}$}constant peculiar velocity, and $\langle \estMfive/\Mx\rangle = 0.86 \pm 0.13$ for the case of 900 \mbox{km\,s$^{-1}$}.  This is for the CILC, but the results are essentially the same for the MMF.  Note that the uncertainties on the sample means are unchanged, because they are calculated from the sample size and the individual measurement error, the latter determined by the unchanged CMB and noise properties.  

There is no evidence of bias in the sample mean at the lower peculiar velocity of 300 \mbox{km\,s$^{-1}$}, while a bias of 14\% appears at the higher value of 900 \mbox{km\,s$^{-1}$}.  In the standard $\Lambda$CDM model, we expect individual cluster peculiar velocities to follow a Gaussian distribution of zero mean and variance $\langle \beta^2 \rangle \approx \left(300\, \mbox{km\,s$^{-1}$}\right)^2$.  Our result for the mean of a set of clusters with this constant velocity is therefore representative of the bias expected of the quadratic term in $\beta$.  The bias term is clearly present, but only important at atypically large peculiar velocities.   

The linear term is also present, causing an increase in the observed dispersion of the individual measurements.  By comparing the dispersion before and after addition of the kSZ effect, we deduce that it contributes 0.14 and 0.31, respectively, for the lower and higher peculiar velocities.  The former value is the more realistic and should be compared to the CMB and noise contribution to the dispersion (individual uncertainties) of 1.06.

\begin{figure}[htb]
\begin{center}
\includegraphics[width=0.5\textwidth]{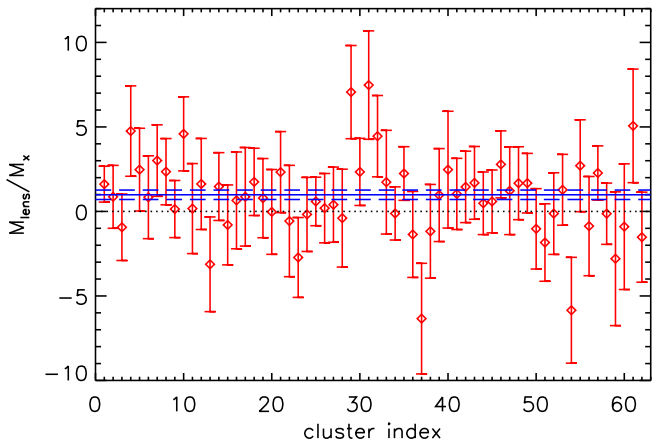}
\end{center}
\caption{Recovered mass for the \Planck\ ESZ-XMM simulation, expressed as in Fig.~\ref{fig:result_2dilc} by the ratio of the CMB-measured mass to the input mass from the X-ray model, $\estMfive/\Mx$.  The tSZ was removed using the CILC, and the blue solid and dashed lines show the sample mean and its $1\sigma$ range:  $\langle \estMfive/\Mx \rangle= 0.99\pm 0.28$.  This simulation includes a random kSZ effect with Gaussian standard deviation of 300 \mbox{km\,s$^{-1}$}.  The CMB and noise realizations are the same as in Fig.~\ref{fig:result_2dilc}, resulting in a similar pattern in the distribution of points, but individual uncertainties are larger (note the change in scale).}
\label{fig:result_xmmesz}
\end{figure}

The overall conclusion is the same for the simulated ESZ-XMM, as shown in Fig.~\ref{fig:result_xmmesz}.  We obtain an unbiased estimate of the sample mass scale: $\langle \estMfive/\Mx \rangle= 0.99\pm 0.28$.  The significance is lower than before because the ESZ-XMM contains a range of cluster masses, all of them smaller than A2163, its most massive member.  Nevertheless, we find that the mass scale can be recovered at the $3\sigma$ level. In this case, truncating the profile at $R_{vir}$ increases the uncertainty from 0.28 to 0.33, leaving the conclusion unchanged. This simulation also included a kSZ effect from random peculiar velocities with a Gaussian standard deviation of 300 \mbox{km\,s$^{-1}$}.  We see no evidence for its impact on the recovered sample mean.

\section{Discussion}
\label{sec:discussion}
Our results show that the proposed method can be a practical tool for estimating cluster masses, even with current CMB temperature data.  It offers a new way to constrain cluster scaling relations between total mass and observables such as X-ray luminosity, SZ signal strength or richness.  These relations are central to cosmological and large-scale structure analyses of cluster catalogues.

The method complements others for measuring cluster mass. Like gravitational shear, it directly probes total mass without assumptions about the state of any particular cluster component; this is strictly true for the CILC method, although we note that there does remain some modeling uncertainty when applying the MMF component separation, as discussed above.
The method's particular strengths are that it can be used to much higher redshifts, using the CMB as a source plane, and that it is sensitive to the convergence field, rather than its gradient like the shear.  Noting this latter difference, several authors have recently pointed out the value of combined CMB and shear analyses in the more general context \citep{hand2013,das_lensing2013}.  

A critical capability of the proposed procedure is to provide accurate measurements and their uncertainties in the low signal-to-noise regime.  This enables statistical analyses that permit practical application to existing CMB temperature datasets.

We achieve it through the lensing reconstruction that furnishes well-defined noise properties for use by the filter.  The noise arises not just from the instrument, but also from the CMB itself, because the lensing signature is a correlation between CMB anisotropy modes behind the cluster, modes that we do not a priori know.  Instead, we rely on the power spectrum of the primary CMB (and its Gaussianity) to tell us what they are on average and the dispersion about that average.  In many ways, this is simpler than trying to determine the CMB gradient around each individual cluster, as needed when working directly with the secondary temperature anisotropy generated by lensing.

Figure~\ref{fig:Mlim} summarizes the ideal statistical power of various experiments to measure cluster mass, showing the standard deviation of filter mass determinations as a function of cluster redshift.  It is ideal because it only accounts for noise from the CMB and the instrument. 
The curves are extremely flat in redshift, a reflection of the broad lensing kernel to the source plane of the CMB.  

Comparing the space experiments (the red curves), we see that \Planck\ is dominated by instrumental noise, since it has the same angular resolution as COrE, which performs much better with its lower noise level, while PRISM gains further by incorporating more modes in reaching to smaller scales.
Even at their higher angular resolution, the ground-based experiments are not dominated by the CMB, as can be seen from the fact that the mass filter noise continues to decrease with decreasing instrumental noise.  

We examined the impact of the tSZ and kSZ signals using simulations of \Planck-like observations of massive clusters.  Our main result is that the tSZ must be accurately removed to avoid biasing the mass estimations.  A standard ILC is not sufficient.  We applied two methods, both of which proved satisfactory.  Based on the ILC, their key additional characteristic is that they actively subtract or nullify the tSZ.  The success of the CILC is encouraging, because it does not rely on any adopted profile for the tSZ.  Our result emphasizes the importance of broad frequency coverage in experimental design to enable adequate component separation.

The kSZ cannot be removed through such spectral separation methods and it remains in the CMB maps used to reconstruct the lensing potential.  It impacts the final result by adding a source of measurement noise (term linear in $\beta$) and a bias (term in $\beta^2$).  In our simulations, we found evidence for both terms.  Fortunately, their influence is small for the expected distribution of cluster peculiar velocity.

These latter conclusions only apply in the context of our simulations of \Planck-like CMB observations of massive clusters.  Understanding these details in other experimental setups, e.g., ground-based instruments, would require dedicated simulations.  The same applies to study of \Planck-like observations of large cluster samples including lower mass systems.  

Additional limitations of the simulations presented here include lack of other foreground sources, such as diffuse Galactic emission and extragalactic sources.  Since we are using component separation techniques based on the ILC, we assume for this preliminary study that they adequately remove these foregrounds.  Our clusters are also simulated in isolation and modeled by the same spherically symmetric profiles used in our filters (tSZ and mass filters)\footnote{Recall the above statement that the CILC does not, however, rely on an adopted template.}.  More realistic simulations would employ variations in model profiles and numerical simulations of clusters in their cosmological context to evaluate the effects of local structure around the clusters, as well as that of uncorrelated large-scale structure along the line-of-sight that contribute to the noise term.  

Despite these limitations, our simulations are sufficient to demonstrate the potential of the \Planck\ 2013 dataset to detect the mass scale of the ESZ-XMM.  Our simulation of this observation of 62 clusters is summarized in Fig.~\ref{fig:result_xmmesz}.  The conclusion is that we should be able to detect the mass scale of this catalog to slightly more than $3\sigma$.  A separate paper presents our analysis on the real \Planck\ dataset.

\section{Conclusion}
\label{sec:conclusion}
We propose a method to measure galaxy cluster masses using CMB lensing and demonstrate that it can be practically applied to existing datasets (e.g., ACT, \Planck, SPT) in statistical analyses of cluster samples.  The strength of the approach draws from its ability to provide viable mass estimates and uncertainties even in low signal-to-noise regimes, thereby enabling straightforward statistical analyses of systems well below individual detection.  

Accurate removal of the tSZ is important and achievable, as we demonstrate by application of component separation methods that actively subtract or nullify it.  The implication is that experimental design must allow for sufficient spectral coverage to enable effective separation methods.  

Using a simulation of the \Planck\ ESZ-XMM sample, we conclude that it would be possible to determine the mass scale of this set of 62 clusters to $3\sigma$ significance (CMB and instrumental noise only).  In a companion paper, we present a first application of our method to the \Planck\ data on the actual ESZ-XMM.

The method presented here uses only temperature data in the lensing reconstruction.  Future work will extend it to CMB polarization data.  Our preliminary study here opens the way to numerous research avenues targeting additional issues related to foregrounds and large-scale structure, and calls for detailed studies dedicated to specific experimental campaigns.

Lensing of the CMB opens a new and independent avenue for studying cluster masses, an important complement to other techniques, such as weak gravitational lensing of background galaxies.  In fact, CMB lensing offers the possibility of calibrating large cluster samples now while we await large area galaxy lensing surveys, such as the Dark Energy Survey, the Large Synoptic Survey Telescope, and the {\em Euclid} and WFIRST space missions.  And it will remain the more efficient way to measure cluster masses at high redshifts, where the source galaxy population rapidly declines in imaging surveys. 

\begin{acknowledgements}
The authors would like to thank the anonymous referee for useful comments which helped to clarify some important aspects of this work.  A portion of the research described in this paper was carried out at the Jet Propulsion Laboratory, California Institute of Technology, under a contract with the National Aeronautics and Space Administration.
\end{acknowledgements}

\bibliographystyle{aa}
\bibliography{cmblens}

\end{document}